\documentclass[12pt]{article}
 \pdfoutput=1
\newcommand{\mb}{\mathbf }

\newcommand{\pr}{^\prime}
\newcommand{\ca}{{Carath\'{e}odory}\,\,}
\usepackage{graphicx}
\usepackage{epstopdf}
\DeclareGraphicsExtensions{.jpg,.pdf}
\usepackage[numbers]{natbib}
\usepackage{rotating}
\usepackage{graphicx}
\usepackage{amsmath}
\usepackage{amssymb}
\begin{document}
\title{Second Law Considerations in Fourier Heat Conduction of a Lattice Chain in Relation to Intermolecular Potentials}
\author{Christopher G. Jesudason 
\thanks{Emails: jesu@um.edu.my; chris\_guna@yahoo.com  }\\
 {\normalsize   Department of  Chemistry and Center for Theoretical and Computation Physics  } \\ 
 {\normalsize University of Malaya}\\
  {\normalsize 50603 Kuala Lumpur, Malaysia}\\
  {\footnotesize currently on  leave at} \\ 
  {\footnotesize Department of Chemical and Geological Sciences,}\\
  {\footnotesize University of Cagliari, Sardinia, Italy.}
  }%\da

% Hint: \title{what ever}, \author{who care} and \date{when ever} could stand 
% before or after the \begin{document} command 
% BUT the \maketitle command MUST come AFTER the \begin{document} command! 
%\begin{document}
\date{}
\maketitle

% Title portion

%\author[aff1,aff2]{\corref{cor1}}
%\eaddress[url]{https://umexpert.um.edu.my/jesu}
%\author[aff2,aff3]{Author's Name}
%\eaddress{anotherauthor@thisaddress.yyy}

%\affil[aff1]{Chemistry Department and  Center for Theoretical and Computation Physics, Science Faculty, University of Malaya, Pantai Valley, 50603 Kuala Lumpur, Malaysia.}
%\affil[aff2]{Currently on sabbatical leave at Department of Chemical and Geological Sciences, University of
%Cagliari, Cagliari, Sardinia, Italy. \corref{cor2}}
%\affil[aff3]{You would list an author's second affiliation here.}
%\corresp[cor1]{Corresponding author: jesu@um.edu.my;%\,chrysostomg@gmail.com}
%\corresp[cor2]{Corresponding author: jesu@um.edu.my}

%\author[aff1]{Author's Name\corref{cor1}}
%\eaddress[url]{http://www.aip.org}
%\author[aff2,aff3]{Author's Name}
%\eaddress{anotherauthor@thisaddress.yyy}

%\affil[aff1]{Replace this text with an author's affiliation (use complete addresses). Note the use of superscript ``a)'' to indicate the author's e-mail address below. Use b), c), etc. to indicate e-mail addresses for more than 1 author.}
%\affil[aff2]{Additional affiliations should be indicated by superscript numbers 2, 3, etc. as shown above.}
%\affil[aff3]{You would list an author's second affiliation here.}
%\corresp[cor1]{Corresponding author: your@emailaddress.xxx}

%\maketitle

\begin{abstract}
Two aspects of conductive heat are focused here (i) the nature of conductive heat, defined as that form of energy that  is transferred as a result of a temperature difference and (ii) the nature of the intermolecular potentials that induces both thermal energy flow  and the temperature profile at the steady state for a 1-D lattice chain. It is found that the standard presuppositions of people like Benofy and  Quay (BQ) following Joseph Fourier do not obtain for at least a certain specified regime of intermolecular potential parameters related to harmonic  (quadratic) potentials for nearest neighbor interactions. For  these harmonic potentials, it appears from the simulation results that steady state solutions exist utilizing non-synthetic thermostats that couple not just  the two particles at the extreme ends of the lattice chain, but to a control volume of $N$ particles located at either ends of the chain that does not accord with  the unique analytical solutions that obtains for single particle thermostatting at the ends of the lattice with a different thermostatting algorithm that utilizes coupling coefficients. If the method used here is considered a more "realistic" or  feasible model of the physical reality, then a re-evaluation of some aspects of the  standard theoretical methodology  is warranted  since  the standard model  solution profile  does not accord with the simulation temperature profile determined here for this related model. We also note  that the sinusoidal  temperature profile  generated suggests  that thermal integrated circuits  with several thermal P-N junctions may be constructed, opening a way  to create more complex  thermal transistor circuits. A stationary principle is proposed for regions that violate the Fourier principle $\mathbf{J_q.}\nabla T \le 0 $, where $\mathbf{J_q}$ is the heat current vector and $T$ the temperature.
% The AIP Proceedings article template has many predefined paragraph styles %for you to use/apply as you write your paper. To format your abstract, use the \LaTeX template style: {\itshape Abstract.} Each paper must include an abstract. Begin the abstract with the word ``Abstract'' %followed by a period in bold font, and then continue with a normal 9 point %font.
\end{abstract}

% Head 1
\section{INTRODUCTION}\label{sec:1}
Whenever the Fourier law obtains, (here confined to the linear first order version ) $\mathbf{J_q}=\kappa \nabla T(\mathbf{r})$ where  $\mathbf{J_q}$ is the heat current vector, $\kappa$ the thermal conductivity and  $ T(\mathbf{r})$ the temperature at coordinate $\mathbf{r}$, then Fourier maintained that \cite[Sec.III, no. 57-64, pp.41-45]{four1} (a) net heat energy flow cannot occur in the absence of a temperature gradient,  and (b) net heat flow occurs from hot to cold temperature regions that are connected if a temperature gradient exists. With the implication of  local  behavior, his postulates (a) and (b) are taken to imply 
\begin{equation}\label{e:1}
\mathbf{J_q.}\nabla T \le 0,
\end{equation}
where (a) and (b)  taken together  refer to the Fourier (\textbf{F}) principle in (\ref{e:1}).  Fourier and his followers claim that conductive heat is local in nature (within the limits of molecular volumes and particle interaction times) with (\ref{e:1}) obtaining where 
 Benofy and Quay \cite[p.11]{bq}  following Fourier have argued that the Fourier law is essentially local in nature, where  whenever a temperature gradient is present, there can be a flow of heat but  there cannot be such conductive  heat transfer in the absence of a thermal gradient.  BQ also argue that  the Second law statements of Kelvin and Clausius are global, so that with compensation, there can be transfer of heat from cold to hot, but never by conduction \cite[p.10, par. 2-3]{bq}. The fundamental  definition of heat, according to some authorities,  on the other hand is that form of energy that traverses  a boundary as a result of a temperature difference  (\cite[p.73]{zemansky1}, \cite[p.229]{cara1},\cite{cara2}).  Further a direction of traverse is also implied. Carath\'{e}odory defines heat  ( \cite[J. Kestin ed., Introduction, p.229]{cara1} as follows:  {\em"`Furthermore, when two bodies of different temperatures are brought into contact, heat always passes from the hotter to the colder, and never in the reverse direction."'} In passing, the  more restricted previous work \cite{cgj30} identifies Fourier  conductive heat  transfer as thermodynamical "heat" and  showed that this heat actually conforms to a Carnot optimized trajectory. From these definitions, one can surmise that a contradiction to (\ref{e:1}) implies that conductive heat is not only local within the aforementioned limits, but could involve some type of "optimized"  trajectory where global principles apply even within  a localized region. This  is another result which is  postulated in what follows based on the simulation results.

 \section{SYSTEM DESCRIPTION AND RESULTS}
 A work of considerable prescience that has proved influential to the entire field of low-dimensional heat transfer is that by Rieder, Lebowitz and Lieb (RLL) \cite{rieder1} where for the first time a unique solution to the harmonic lattice chain was  provided (officially) under restricted thermostatting conditions. There is mention of an unpublished work of  O.Bills as having foundation significance in their derivation \cite[their ref. 6]{rieder1}, as well as other official works referenced. Here the steady state temperature profile $T(j)$ and total current density $J(\lambda, \omega)$ \cite[eq.(4.6)]{rieder1}  are some of the quantities derived for a  harmonic interparticle potential 1-D lattice of equal masses; $j$ is the particle index $1$ to $N$, counting from  left $L$ to right $R$ . The model  is  at times vaguely described. RLL speaks of "pistons" of systems interacting with heat baths and then later revert to the two endpoint particles that are thermostated, where the Hamiltonian  on the other hand is  of their  standard form \cite[their eq.2.1]{rieder1} below in (\ref{e:2}), 
 \begin{equation}\label{e:2}
 H=\frac{1}{2}\sum_{i=N}^{2N}x_i^2 + \frac{1}{2}\sum_{i,j=1}^{N}\Phi_{ij}x_ix_j  \qquad N=s\mathcal{N}
 \end{equation}
 where $\Phi$ is the force matrix,  the $x$'s are position ($i=1,N$)-momentum ($i=N,2N$) coordinates with  $\mathcal{N}$ being the number of particles of dimension $s$ each.  The input and output thermal energy channels  are at the first particle 1 at the left at temperature $T_l$  and particle $\mathcal{N}$ on the right of the lattice at temperature $T_N$. There is derived \cite[eq.(3.1)]{rieder1} a heat reservoir interaction parameter $\lambda,\, \lambda_1=\lambda_N=\lambda$ where the heat transfer rate $J(\lambda,\omega)$ is given by 
\begin{equation}\label{e:3}
J(\lambda,\omega)=\left\{ 
\begin{array}{rl}
\frac{1}{2}(\omega^2/\lambda)k(T_1-T_N), \qquad \lambda \gg \omega \\
\frac{1}{2}\lambda k(T_1-T_N), \qquad \lambda  \ll\omega
\end{array}\right.
\end{equation} 
Here, the  energy transfer rate is proportional to  $(T_1-T_N)$  and not on any gradient with respect to distance or particle index $j$.  
 \subsection{Results}
Relative to the assumptions, the solutions are shown to be unique \cite[p.1077,last par.,1st column]{rieder1}. A sketch of the solution for the temperature profile is given in Fig.(\ref{fig:1}).  The plateau in the middle portion is not constant but close to $T=(T_1 +T_N)/2$ . Of importance is that the curve at $j=2$ falls below the mean temperature, and if the Fourier parametrization $\mathbf{J_q.}\nabla T \le 0 $ is used, where $\kappa(j)> 0$ (a fundamental kinetic assumption in thermodynamics), then 
 the \textbf{F} principle fails along this temperature profile segment. The plateau portion is widely quoted in numerical and theoretical studies, over the last half century, where the harmonic potential yields "ballistic trajectories" \cite[p.361]{tejal1}.
 The profile presented in Fig.(\ref{fig:1}) is taken to imply the failure of the Fourier law, for the plateau region suggests $\kappa(j)\rightarrow \infty$ there, whereas portions of the end-point regions suggests $\kappa(j)\leq 0$ . Dhar has opined that \cite[p. 459]{dhar2} Fourier's law is "probably not valid in one- and  two-dimensional systems, except when the system is attached to an external substrate potential." A presumed well behaved system would therefore have Hamiltonian form $H$  \cite[eq. (3)]{dhar2} where
 \begin{equation}\label{e:4}
 H= \sum_{i=1}^N \left\lbrack \frac{p_l^2}{2m_l} +V(x_l) \right\rbrack + \sum_{i=1}^{N-1}U(x_l-x_{l+1})
 \end{equation}
and $V(x_l)$ is the position coordinate  dependent site or substrate potential and $U(x_l-x_{l+1})$ the interparticle nearest neigbor interactions, where the $x$'s  are the spatial coordinates  relative to the equilibrium position. Shah et al. \cite[p.361]{tejal1} on the other hand seem to indicate from their extensive numerical work  that the "general outcome of these studies  is that anharmonicity is the necessary ingredient for the formation of a temperature gradient". Taking this remark as an observation, we carry out simulations where the site potential in (\ref{e:4}), $V(x_l)=0$ (contradicting Dhar but affirming  Shah et al.)   and consider the anharmonic portion $U$ written as 
 
 \begin{equation}\label{e:5}
 U(x_{i-1},x_i)= k_h\frac{(x_i-x_{i-1})^2}{2} +b_h\frac{(x_i-x_{i-1})^4}{4}
 \end{equation}
 which is the FPU-$\beta$ model  for the anharmonic  contribution to the  interparticle potentials \cite[eq.(2)]{tejal1}. Clearly the potentials used are arbitrary. We conform to the standard models of potentials representing the  most basic representative potentials  used in simulation  and  theory so that comparisons may be made.
 
\begin{figure}[htbp]
  \centerline{\includegraphics[width=18cm]{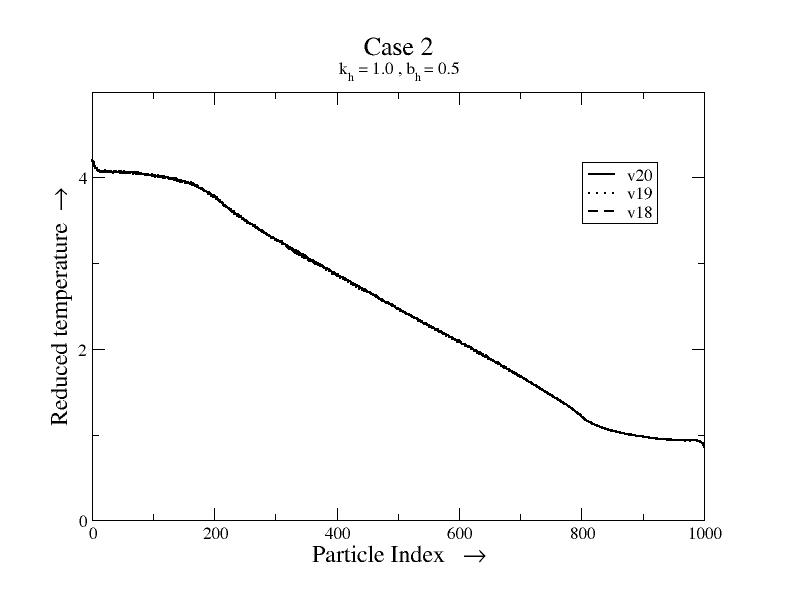}}
  \caption{Case 2: A system  where the \textbf{F} principle obtains everywhere along the lattice chain.}
  \label{fig:3}
\end{figure}

The converged values used for the  $k_h$ and $b_h$  variables  in various runs are as follows:
\begin{enumerate}
\item Case 1, $k_h=1.0, b_h = 0.0$
\item Case 2, $k_h=1.0, b_h = 0.5$
\item Case 3, $k_h=593.355, b_h = 0.0$
\item Case 4, $k_h=593.355, b_h = 0.5$.
\end{enumerate}
The value $k_h=1.0,b_h=0.5, T_L=4.0, T_R=1.0$ was chosen from previous work \cite{cgj30} based on the $b_h$ value from \cite{tejal1} without the site potential $V(x_l)=0$ and the temperature particle index  profile is given  in  Fig.(\ref{fig:3}) as Case 2. 

The Case 2 results  partially verifies Shah et al. (op. cit.) concerning anharmonicity  contributions allowing for Fourier's law to hold without a site potential. The data from the Case 2 system was used to construct a theory of recoverable  conversions  for heat/work energy transitions along  the so-called  recoverable trajectory $\delta\mathcal{S}=0$ \cite{cgj30} as exemplified by Fourier heat conduction where the heat  flow direction  was consonant with the \textbf{F} principle for that particular system throughout. In current terminology, the term "ballistic" terminology applies to solutions  given for instance by the RLL system where  the \textbf{F} principle breaks down \cite[1. Introduction]{tejal1} and where presumably, because of the flat curve, the  conductivity would tend to infinity.  Dhar opines   that the Green's function approach can explain harmonic systems but does not produce any temperature profile in his examples,  and also states \cite[p.460]{dhar2} "..In the present  context ballistic transport means that phonon-phonon interactions can be neglected". How one can describe phonons in classical simulations  is one area that is not so apparent. Hu et al. \cite[p.2994]{bam1}
write : "If the lattice is absent, and the interparticle potential is
harmonic, then no phonon-phonon interaction exists; thus the
heat transfer would take place at the speed of sound and the
thermal conductivity would be infinite, as pointed out by
Debye in 1914. However, if we add a dissipative term to the
harmonic oscillator chain, then we could obtain the Fourier
heat law, even though we do not have a lattice." The originators of the RLL theory \cite{rieder1} claim "no explanation is offered for this paradoxical result" which could refer to the entire temperature profile, or to the portions that violate the \textbf{F} principle. There seems to  be a diversity of opinion in the absence of comprehensively stated theories \cite{lebo2}. 
 
 \subsubsection{Brief Description of Simulation System}
 In all the simulations conducted, the chain length was 1000 particles, of  unit mass. The first $N_s =200$ particles were thermostatted at $T_L=4.0$ and the last 200 at $T_R=1.0$ This method  is to be contrasted to those  where the particles  at the extreme ends  are  thermostatted by various unspecified or synthetic algorithms \cite{vach1} such as the "reversible" 
 Nos\'{e}-Hoover thermostat when it has been proven that  time reversible motion as  utilized in mathematical physics is often misused and misconstrued \cite{cgj4, cgj5, cgj16}. Since several particles are thermostatted, this system differs from the standard RLL and allied models where only the end-point particles are thermostatted and where the algorithm for  thermostatting differs.  The thermostatting method used here is non-synthetic,  where  primes denote the state after the thermostatic move, where we scale  the velocities according to $\mb{\dot{q}\pr_i}=(1+\alpha_I)\mb{\dot{q}}_i+\mb{\beta_I}$, with ${\alpha_I}$  and $\mb{\beta_I}$ being the  parameters to be determined.   If $\mathbf{P}$ is the total momentum  of a control  volume or region, (denoted L and R in this case for the two ends) , then conservation of momentum implies $\Delta \mathbf{P_I} = \mathbf{P^\prime_I} -\mathbf{P_I}=0, \,  \mathbf{I}\in \{L,R\}$. Defining $\mathbf{V_I}=\sum_{i=1}^{N_s}\mathbf{q_i},\,\,W_I=\sum_{i=1}^{N_s}\mathbf{\dot{q}^2_i}$, then to set the temperature we write $W^\prime_i=\frac{3N_skT}{m_I}$ ($m_I$ being the mass of each particle in the control volume)  and we solve the following  equations
\begin{eqnarray}\label{e:6}
\mb{V^\prime}&=&(1+\alpha_I)+ N_s\mb{\beta_I} \\
W^\prime_I &=& (1+{\alpha_I})^2W_I + 2(1+\alpha_I)\mb{\beta_I}\cdot\mb{V_I}+N_s\mb{\beta_I}\cdot\mb{\beta_I}
\end{eqnarray} 
  to determine the scaling parameters $\mb{\beta_I}$ and $\alpha_I$. Currently, no coupling parameters  are used for these thermostats, as opposed to the more synthetic methods, where in these more conventional descriptions, they are important variables for non-equilibrium phenomena as the rate of heat transfer is dependent on the value of these parameters \cite[p.467]{dhar2}.  In standard kinetic theory, the rate of heat transfer \cite[Chap. 15, p.583]{path2} are determined by the kinetic coefficients and the gradients of the thermodynamic potentials or variables, and many phenomenological laws, such as the Fourier heat conduction laws conform to this structure where the kinetic coefficients are dependent on the thermodynamical variables only. On the other hand,  Lepri et al. \cite[Sec 3.3]{lep2} give quantitative  values of how   energy transfer rates vary  with the microscopic coupling values, which is not a feature of conventional theories. 	 Clearly these factors  are very challenging  issue as noted by prominent workers \cite{lebo2} and so far it is not clear whether a comprehensive treatment has been made to remove ambiguity in terms of the actual energy transfer rates  so essential for characterizing these systems.  Dhar   mentions the need for calibration \cite[p.467]{dhar2}. For the MD algorithm here, unlike the use of standard Verlet algorithm for previous studies (e.g. \cite{cgj30}),  the modified  5 stage 4th order method of Calvo and Sanz-Serna \cite{chan1} tested in  reference \citep{gray2} was utilized. The parameters  $(\tau, a_k, b_k)$ for   for this symplectic algorithm was taken from \cite[Table 2]{gray2} where the pseudo-code  for the iterations $(k=1,M, M=5)$ are:
   \begin{eqnarray}\label{e:7}
   \mb{p}^{(k)}&=&\mb{p}^{(k-1)} + b_k\tau\mb{F}(\mb{q}^{(k-1)})\\
   \mb{q}^{(k)}&=&\mb{q}^{(k-1)} + a_k\tau\mb{G}(\mb{p}^{(k)})
   \end{eqnarray}
 where $\mb{F(q)}=-\partial V(\mb{q})/\partial \mb{q}$, $\mb{G(p)}=-\partial T(\mb{p})/\partial \mb{p}$.
 The MD runs each time were $1\times 10^{9}=1\mbox{B}$ time steps after an initial relaxational run of 10M ($1\mbox{M}=10^{-6}$) time steps where the coordinates from the previous run are used for the subsequent one at the commencement of any particular MD run. The results presented in the graphs are typically for  3 runs about  the 17-21th  runs. In the figure legends, $vn$ represents the data for the $n^{th}$ run.  The statistics for the energy transfer were averaged over  30 dumps where each dump sampled 33M steps. The reduced time increment $\delta t$ was $\delta t=0.001$. 
 
 \subsubsection{Discussion}
 
 \begin{figure}[htbp]
  \centerline{\includegraphics[width=18cm]{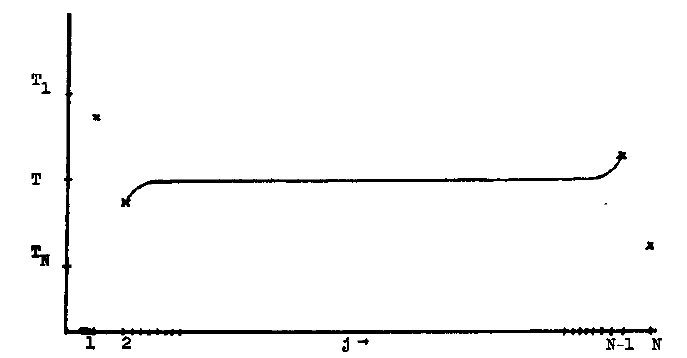}}
  \caption{Sketch of the RLL solution \cite{rieder1}  where the \textbf{F} principle is violated at the ends of the lattice chain and $\kappa(j)\rightarrow \infty $ at the plateau region; $j$ is the particle index and $T$ a scaled temperature.}
  \label{fig:1}
\end{figure}

\begin{figure}[htbp]
  \centerline{\includegraphics[width=18cm]{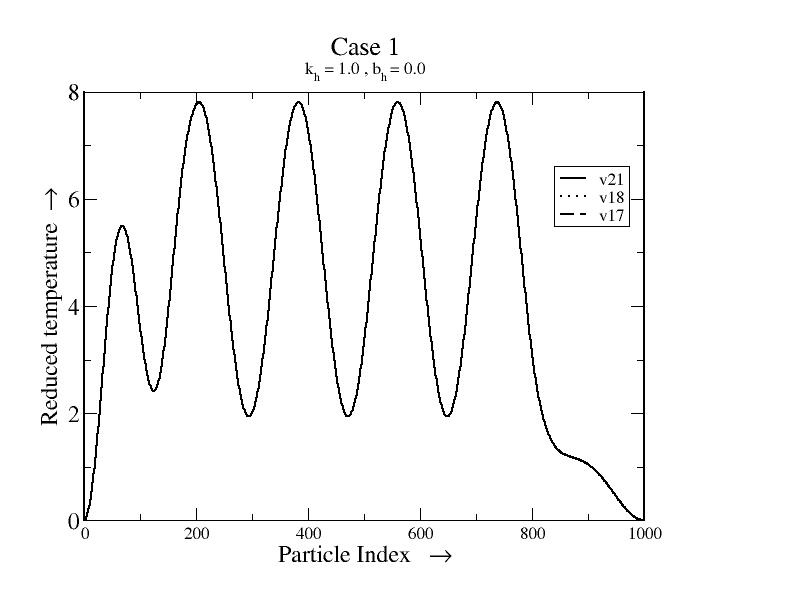}}
  \caption{Case 1: The standard case of of a harmonic lattice chain with clear violation of the \textbf{F} principle along positive temperature gradients. The $k_h=1.0$ value is typical but the system differs from that of  RLL in that the particles in the R and L control volume are thermostatted with 200 particles each, whereas RLL have only one particle each in the control volumes.}
  \label{fig:2}
\end{figure}

 We note that there is little resemblance between the Case 1 profile (Fig.(\ref{fig:2}) to the RLL profile (Fig.(\ref{fig:1})) which is said to be unique, and independent of the temperature difference  between the reservoirs and the absolute temperatures of the reservoirs where in Case 1, the harmonic potential parameter $k_h=1.0$ with no anharmonicity contribution with $b_h=0.0$. We note that RLL used the Liouville equation in conjunction with the Hamiltonian. It was pointed out that the Liouville equation could not in general obtain as   a mathematical truth for systems, although it and the quantum version is the basis for describing systems and might be considered good approximations as a result of their utility \cite[see refs. 66 and 67]{cgj30}. 
 The RLL system and the one studied here are not equivalent since we use a control volume of 200 particles at the R and L ends of the lattice chain for thermostatting using a non-synthetic algorithm with no coupling coefficients. 
 Furthermore, assuming RLL uniqueness to their solution, and that the number of particles thermostatted do not matter, then perhaps the method of coupling of the reservoirs to the particles  play a role in leading to the solution depicted in Fig.(\ref{fig:1}). Another possibility is that these coupling mechanisms may be dependent  on the number of particles thermostatted, and  the temperature and temperature differences. If indeed non-synthetic thermostatting of regions involving multiple particles are deemed to be  reasonable representations of  physically realizable systems, then a re-evaluation of work over the last 1/2 century in theoretical heat transfer is warranted  to incorporate these added features. 
 
 \begin{figure}[htbp]
  \centerline{\includegraphics[width=18cm]{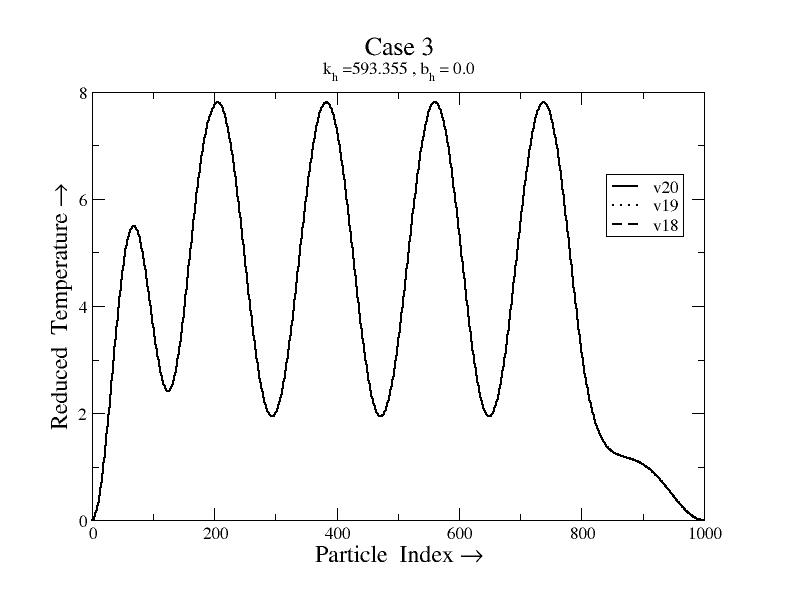}}
  \caption{Case 3: Harmonic lattice chain model with a large $k_h= 593.355$  value. The sinusoidal-like curve has the same temperature-particle index profile  as for Case 1. }
  \label{fig:4}
\end{figure}

\begin{figure}[htbp]
  \centerline{\includegraphics[width=18cm]{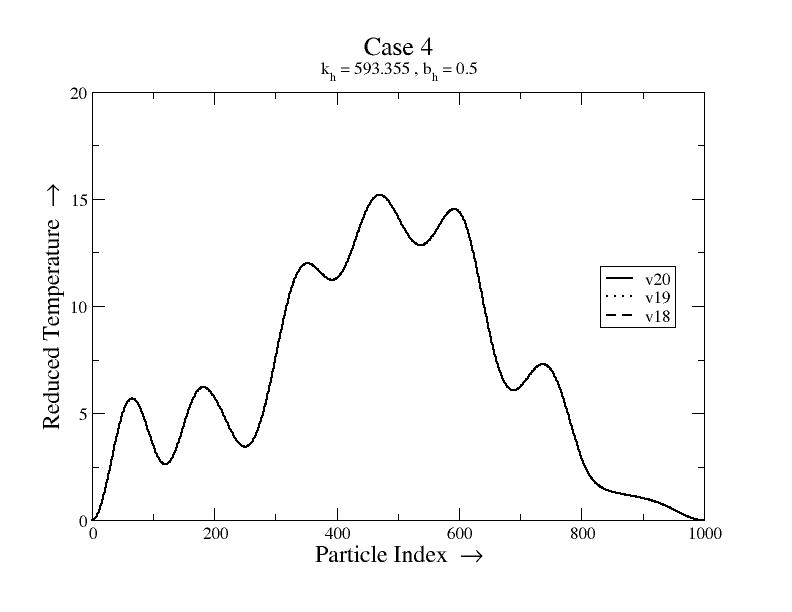}}
  \caption{Case 4: A harmonic lattice chain moderated by an anharmonic contribution but where nevertheless the \textbf{F} principle is violated in some regions along the chain when $k_h$ is  relatively too large.}
  \label{fig:5}
  \end{figure}
 
 The next deduction in this work is related to the problem of the anharmonicity contribution to the potential. Is it a necessary  and sufficient condition to ensure that the \textbf{F} principle  and in particular, the Fourier law obtains?
 Comparing Fig.(\ref{fig:4}) for Case 3  and Fig.(\ref{fig:5}) for  Case 4 allows for a deduction. As a side note,  the (large)  value of $k_h=593.355$ was estimated very approximately by assuming the harmonic potential for the element tungsten (W) with lattice constant 316.52 pm, BCC structure, bulk modulus 310 GPa, where we set the reduced temperature $T^\ast=1\Rightarrow 300 K$. There is nothing else remarkable about this. The anharmonicity constant is the same  $b_h=0.5$  for all simulations  whenever $b_h\neq 0$, where the ratio  $b_h/k_h=8.42\times 10^{-4}\, \mbox{and} \, 0.5$ respectively for Case 4  and Case 2. The purely harmonic Case 1 profile has  5  well defined peaks for $k_h=1$ and Case 3 with the harmonic constant  $k_h=593.355$ has  remarkably  a nearly exact   profile with exact coincidence  of the temperature-particle index  graph,  but with radically different heat transfer rates ($\sim .8388/\mbox{unit time}$) for Case 3 as opposed to ($\sim .3443/\mbox{unit time}$) for Case 1 when the input coordinates for Case 3 were derived from the output for Case 1 at an earlier stage prior to relaxation to a new steady state. We note also the well developed  curves of Case 1 and 3  seem to indicate the formation of quasi-mechanical standing "thermal waves" despite a net dissipative  transfer of heat  from hot to cold reservoirs at different rates. The introduction of anharmonicity ($b_h\neq 0$) with the same reservoir algorithm  and harmonic coupling coefficient $k_h$  value destroys or smoothes out the standing wave pattern, and further another peak (6 peaks) are added with a heat transfer rate of $\sim ..81257/\mbox{unit time}$  for Case 4.
 We therefore conclude that anharmonicity is a necessary but not sufficient condition for the Fourier law to obtain. Indeed the ratio of the force field parameters $(b_h/k_h)$ seems to determine whether the Fourier law is obeyed or not. In addition, whether absolute magnitudes of $(k_h,b_h)$ are also featured in the criterion of force field ratios  is not known at the present time.
  \subsubsection{A Fundamental    Hypothesis Concerning Heat Transfer}
  When the theory of "recoverable transitions" \cite{cgj14} was applied to  Fourier conduction \cite{cgj30}, the current data was clearly unavailable and the RLL result  seemed  to be based on several presuppositions, with the current data not having any resemblance to the RLL construct. The current data  indicates (as with the RLL result at the ends of the lattice chain) that "heat" is flowing   in the direction  of a temperature  gradient  which would invalidate the \textbf{F} principle and that form of energy as heat transfer  in conventional thermodynamical definition. Whilst  the current simulation data does not contradict or invalidate the theory  developed in \cite{cgj30} for a system that complied with the \textbf{F} principle everywhere, we postulate  that the same theory can explain the transfer of energy  along  a positive   temperature gradient, and which  can still be considered as thermal energy (extending the concept of heat as defined by \ca and the definition of heat in thermodynamics) if we simply state that even in regions where the Fourier law is apparently violated , the entropy change along the trajectory is invariant and  may be written 
  \begin{equation}\label{e:8}
  \delta\mathcal{S}\bigg|_{Traj}=0.
  \end{equation}
  
  \section{CONCLUSION}
  The data presented indicates that the RLL interpretation, remarkable as a first attempt in describing anomalous  heat diffusion or Fourier conduction  that has rightly influenced  nearly all subsequent  work  over the last half century  is  probably a  model that could be augmented by a more flexible set of conditions including the nature of thermostatting in terms of coupling mechanisms, the number of particles in the control volumes of the thermostatted particles, and the dynamical equations. This is a major project.   The remarkable quasi-mechanical sinusoidal curves  of the steady state profile for harmonic interparticle potentials imply that  the peaks in those graphs, as well as the troughs  can be coupled to other lattice chains to produce more complex integrated thermal circuits than is currently being investigated \cite{vach1}. We showed  that anharmonicity is a necessary  but not sufficient condition  for  traditional Fourier heat conduction  mechanisms to apply.  We also present a hypothesis that even for harmonic potential lattices, heat flow can occur  along a temperature gradient, in accordance with  recoverable transition theory \cite{cgj14}   as expressed  in (\ref{e:8}), which generalizes the nature of heat and its direction of flow in a temperature gradient as defined in the First and Second laws of thermodynamics.

% Acknowledgement
\section{ACKNOWLEDGMENTS}
University of Malaya is thanked for funding this research  with the UMRG grant RG293-14AFR and for partial funding of the conference.I am very grateful to  Francesa Mocci (Dept. of Chemical and Geological Sciences, Univ  of  Cagliari, 09042 Monserrato, Sardinia, Italy) for arranging a collaboration and sabbatical research placement during the time this research was conducted.  I thank the organizers, in particular Seenith Sivasundaram  and Eva Klasik for their remarkable efficiency and goodwill  in dealing with conference issues.

% References

%\nocite{*}
\bibliographystyle{unsrt}%
\bibliography{icnpaa16}%

\end{document}